\documentclass[12pt]{article}

\usepackage{amssymb}
\usepackage{latexsym}
\usepackage{epsfig}
\usepackage{citesort}

\def  \btosll    {$b \to s \ell^+ \ell^-$}
\def  \bstoll    {$B_s \to \ell^+ \ell^-$}
\def  \bstomm    {$B_s \to \mu^+ \mu^-$}
\def  \bstollg   {$B_s \to \ell^+ \ell^- \gamma$}
\def  \bstommg   {$B_s \to \mu^+ \mu^- \gamma$}
\def  \bstottg   {$B_s \to \tau^+ \tau^- \gamma$}

\def  \etal      {{\sl et al.}}

\begin{document}


\begin{titlepage}

\renewcommand{\thefootnote}{\fnsymbol{footnote}}

\vspace*{-0.5truecm}
\begin{flushright}
{\tt hep-ph/0504193 \\
YITP-05-18}
\end{flushright}
\vspace*{0.5truecm}
\begin{center}
{\Large \boldmath \bf Signatures of new Physics in dileptonic $B$-decays}\\
\hspace{10pt}\\
\vspace{2.truecm}
{\bf S. Rai Choudhury$^a$ \footnote{src@physics.du.ac.in}, 
A. S. Cornell$^b$ \footnote{alanc@yukawa.kyoto-u.ac.jp}, 
Naveen Gaur$^a$ \footnote{naveen@physics.du.ac.in} and 
G. C. Joshi$^c$ \footnote{joshi@physics.unimelb.edu.au}}
\vskip .8cm
$^a${\sl Department of Physics \& Astrophysics, University of Delhi, Delhi 110007, India,} \\
\vspace{0.2truecm}
$^b${\sl Yukawa Institute for Theoretical Physics, Kyoto University, Kyoto 606-8502, Japan,} \\
\vspace{0.2truecm}
$^c${\sl School of Physics, University of Melbourne, Parkville, Victoria 3010, Australia}\\
\vspace{2cm}
{\bf Abstract}
\vspace{.2cm}
\begin{quote}
\noindent Leptonic decays of $B$-mesons are theoretically very clean probes for testing the Standard Model (SM) and possible physics beyond it. Amongst the various leptonic decays of the $B$-meson, the pure dileptonic decay $B \to \ell^+ \ell^-$ is very important as this mode is helicity suppressed in the SM but can be substantially enhanced in some of the models beyond the SM, such as supersymmetric (SUSY) theories and the two Higgs doublet model (2HDM). Although the purely dileptonic decay mode is helicity suppressed in the SM its associated mode $B \to \ell^+ \ell^- \gamma$ does not have the same suppression, due to the presence of the $\gamma$ in the final state. In this paper we will also analyse the effects of enhanced $Z$-penguins on these two decay modes.    
\end{quote}

\end{center}
\end{titlepage}

\setcounter{footnote}{0}


\section{Introduction \label{intro}}

\par Experimental data on the decay of the $B$-meson into two pseudo-scalar mesons seems to indicate a ``{\sl puzzle}", in the sense that it points to a sizeable deviation from standard theories, as pointed out in Buras \etal \cite{Buras:2000gc}. To understand the origins of this ``{\sl puzzle}" consider the decay $B \to \pi \pi$, where this decay is reasonably well described within the theoretical framework of an effective Hamiltonian \cite{Buras:2003dj}. However, extensions of these results using $SU(3)$ symmetries for decays $B \to \pi K$ show considerable disagreement with data \cite{Buras:2000gc,Buras:2003dj,Yoshikawa:2003hb}. A possible resolution of the ``$B \to \pi K$" puzzle by Buras \etal \cite{Buras:2000gc} has  aroused considerable interest \cite{Buras:2000gc,Buras:2003dj,Yoshikawa:2003hb}. The explanation is to attribute this deviation to an enhanced $Z$-penguin diagram in addition to this diagram having an additional large phase. This explanation has aroused much interest as it could be the first indication of physics beyond the standard model (SM), where it has been usually accepted that the values of the CKM matrix cannot reproduce such a large phase in the $Z$-penguin diagram.

\par The idea of a strongly enhanced $Z$-penguin contribution is not a new one. It was first carried out by Colangelo and Isidori \cite{Colangelo:1998pm} in relation to the $K \to \pi \nu \bar{\nu}$ and $K \to \pi \ell \bar{\ell}$ decays. Note that Buras and Silvestrini \cite{Buras:1998ed} placed constraints on these contributions in a general class of supersymmetric (SUSY) models. The possibility of nonstandard $Z$-couplings in the context of $b \to s \ell \bar{\ell}$ transitions was studied in \cite{Buchalla:2000sk}. The possible implications of this were studied in some of our earlier works \cite{RaiChoudhury:2004pw}. However, \cite{Buras:2000gc} were the first to relate the possible enhancement of the $Z$-penguins to the nonleptonic decay modes involved in the ``$B \to \pi K$" puzzle and were able to obtain definitive phenomenological values of the magnitude and phase of the $Z$-penguin that is consistent with the $B \to \pi \pi$, $\pi K$ data. The estimates of the magnitude and phase of the $Z$-penguin required to fit the $\pi \pi$ and $\pi K$ data, which have been made in reference \cite{Buras:2003dj}, are purely on a phenomenological basis. On the theoretical side, such an enhancement of the penguin diagrams can be accommodated within SUSY extensions of the SM, the minimal supersymmetric standard model (MSSM) in particular. The flavour rotation of the squarks is different in such theories from the corresponding flavour rotation of the quarks, and this mismatch becomes the source of an additional phase in flavour changing amplitudes. Any attempts to fit the evaluated value of the $Z$-penguin diagrams with theory, however, is hopeless, since the parameters involved in estimating the resultant phase, which are essentially the off-diagonal elements of the squark mass matrix, are not known \cite{Lunghi:1999uk}. Irrespective of this, the occurrence of a phase (other than the CKM phase) in the $Z$-penguin is a signal for a new source of CP violation, and has wider implications.

\par The basic vertex involved in the analysis of Buras \etal \cite{Buras:2000gc} is the $bsZ$ vertex. Processes which feature this vertex have been widely studied in the context of many other semi-leptonic and hadronic processes \cite{Buras:2000gc,Buras:2003dj,Yoshikawa:2003hb,RaiChoudhury:2004pw}. In this work we will address the implications of the large $bsZ$ phenomenological phase to the processes $B \to \ell^- \ell^+$ and $B \to \ell^- \ell^+ \gamma$.

\par Dileptonic decays of $B_s$ are very special rare leptonic decays; the non-radiative mode \bstoll is helicity suppressed in the SM whereas the radiative mode \bstollg does not suffer from such a suppression in the SM. Thus, the later can be comparable to the former despite an extra factor of $\alpha$ (the electromagnetic coupling constant). Recently the CDF and D0 collaborations \cite{dilep_exp1} have reported a very useful bound on the pure dileptonic mode and in the near future we may expect data on both the radiative and non-radiative dileptonic modes. This motivates us to look afresh for the signatures of new physics in these two decays.

\par Within the SM, the pure dileptonic decay is dominated by the $Z$-penguin and box diagrams, both of which are helicity suppressed. It was first recognized by two of us \cite{Choudhury:1998ze} that the situation changes in SUSY theories if the parameter $tan\beta$ is large, and this has been the subject of investigations in a large number of later works \cite{Babu:1999hn,Bobeth:2001sq}. Similar sorts of enhancements are also possible in two Higgs doublet models, as has been emphasized in many earlier works \cite{Skiba:1992mg,Logan:2000iv}. It was also pointed out by Handoko \etal \cite{Handoko:2001bs} that possible measurements of the longitudinal polarizations in the pure dileptonic case will be a useful signature of new physics when SUSY effects with large $tan\beta$ are involved. This polarization is sensitive to the phase of the effective $bsZ$ vertex and hence its measurement can also provide us with insights into new physics.

\par When a photon is emitted in addition to the lepton pair in a purely dileptonic decay, no helicity suppression exists. This process, $B \to \ell^+ \ell^- \gamma$, is thus of special interest among the rare $B$-decays. There have been several investigations of the sensitivity of possible measurables, like forward-backward (FB) asymmetries and polarization asymmetries on new physics \cite{Mohanta:2005gm,RaiChoudhury:2002hf,Kruger:2002gf,Aliev:1996ud,Geng:2000fs,Dincer:2001hu}. Although this process is, theoretically, somewhat less clean as compared to the pure dileptonic mode, in so far as it requires a knowledge of the $B_s \to \gamma$ form factors, this process has a greater number of observables than the pure dileptonic decay. An enhanced $bsZ$ coupling with a phase will affect all these parameters, and motivates us in to calculating the various observables in the radiative dileptonic mode. There is also a second reason for reinvestigating this decay. In our earlier work we used form factors which did not meet certain formal constraints, as established by Kr\"{u}ger and Melikhov \cite{Kruger:2002gf}, who suggested in their paper a recalculation of our results with a new set of form factors. We have also done this here with the inclusion of a non-standard $bsZ$ coupling, as stated above.

\par In the present work we will examine the pure dileptonic decay mode (\bstoll) and the radiative mode (\bstollg) in two different models, namely the two Higgs doublet model (2HDM) and the minimal Supergravity (mSUGRA) model. As such, this paper shall be organized along the following lines. In section 2 we shall present the effective Hamiltonian for the quark level transition \btosll, before specialising to the purely dileptonic decay ($B \to \ell^- \ell^+$) and deriving its matrix element and other observables associated with it. The final subsection of this section of the paper shall analyze the radiative process ($B \to \ell^- \ell^+ \gamma$), deriving expressions for its matrix element, branching ratio and FB asymmetry. The paper shall be concluded in section 3 with our numerical results, discussions and conclusions.


\section{The Effective Hamiltonian \label{section2}}

\par The processes \bstoll and \bstollg are based on the quark process $b \to s \ell^+ \ell^-$ which can be described by the effective Hamiltonian;
\begin{eqnarray}
{\cal H}_{eff} & = & \displaystyle \frac{\alpha G_F}{\sqrt{2} \pi}
V_{tb} V_{ts}^* \Bigg[ - 2 C_7 \frac{m_b}{q^2} \left( \bar{s} i
\sigma_{\mu \nu} q^{\mu} P_R b \right) \bar{\ell} \gamma^{\mu} \ell +
C_9^{eff} \left( \bar{s} \gamma_{\mu} P_L b \right) \bar{\ell}
\gamma^{\mu} \ell   \\   \nonumber  
&& \hspace{1in} + C_{10} \left( \bar{s} \gamma_{\mu} P_L b \right)
\bar{\ell} \gamma^{\mu} \gamma_5 \ell 
+ C_{Q_1} \left( \bar{s} P_R b \right) \bar{\ell}
\ell + C_{Q_2} \left( \bar{s} P_R b \right) \bar{\ell} \gamma_5 \ell
\Bigg] , 
\label{Hamiltonian}
\end{eqnarray}
where $q_{\mu} = (p_+ + p_-)_{\mu}$ and $P_{L/R} = \frac{1}{2} ( 1 \mp \gamma_5 )$. The values of the various Wilson coefficients are given in \cite{Barger:1989fj,Logan:2000iv} for the 2HDM model and \cite{Bobeth:2001sq} in the SUSY extensions of the SM. $C_9^{eff}$ has the value;
$$ C_9^{eff} = C_9 + Y(q^2) , $$
where $Y(q^2)$ has contributions from one loop matrix elements of four quark operators. This, along with short distance contributions, also has long-distance contributions due the $c \bar{c}$ resonant states. These contributions we have incorporated using the prescription given in Kr\"{u}ger and Sehgal \cite{Kruger:1996cv}. Using the above definition of the effective Hamiltonian we will derive expressions for the observables we want to study.


\subsection{The Dileptonic Decay ($B_s \to \ell^+ \ell^-$)\label{dilep:1}}

\par To evaluate \bstoll we need hadronic matrix elements of various currents between $B_s$ and the vacuum. These hadronic matrix elements, using the PCAC ansatz, can be formulated as;
\begin{eqnarray}
\langle 0 | \bar{q} \gamma_{\mu} \gamma_5 b | B(p_B) \rangle & = & - i
f_{B} p_{B \mu} , \nonumber \\
\langle 0 | \bar{q} \gamma_5 b | B(p_B) \rangle & = & \frac{ i f_{B}
m_{B_q}^2}{m_b} , \nonumber \\
\langle 0 | \bar{q} \sigma^{\mu \nu} P_R \gamma_5 b | B(p_B) \rangle &
= & 0 . \label{eq:1:1}
\end{eqnarray}

\par From the effective Hamiltonian given in eqn.(\ref{Hamiltonian}), and using the above definition for the form factors, we can write the matrix element as;
\begin{equation}
{\cal M}  =  i f_{B_q} \frac{G_F \alpha}{\sqrt{2} \pi}
V_{tb} V_{ts}^* \Bigg[ \left( 2 m_{\ell} C_{10} + \frac{m_{B}^2}{m_b}
C_{Q_2} \right) \bar{\ell} \gamma_5 \ell + \left(\frac{m_B^2}{m_b}
C_{Q_1}\right) \bar{\ell} \ell \Bigg] . 
\label{eq:1:2}
\end{equation}

\noindent From the above matrix element we can evaluate the branching ratio as;
\begin{eqnarray}
{\cal B}(B_s \to \ell^+ \ell^- )  
&=&  \frac{ G_F^2 \alpha^2}{64 \pi^3} 
\left| V_{tb} V_{ts}^* 
\right|^2 f_B^2 m_B \sqrt{1 - \frac{4 m_{\ell}^2}{m_B^2}}
\Bigg[ \left| 2 m_{\ell} C_{10} + \frac{m_B^2}{m_b}
C_{Q_2} \right|^2 \nonumber \\
&& + \left( 1 - \frac{4 m_{\ell}^2}{m_B^2} 
\right) \left| \frac{m_B^2}{m_b} C_{Q_1} \right|^2 \Bigg] .
\label{eq:1:3}
\end{eqnarray}

\par In the pure dileptonic ($B_s \to \ell^+ \ell^-$) decay we have only one momenta available in the final state. Hence, we can have only one polarization asymmetry, namely the longitudinal one. This polarization asymmetry provides a direct means for measuring the scalar and pseudoscalar type interactions, which are induced in almost all variants of the 2HDM and SUSY models.

\par Defining the longitudinal polarization asymmetry of the final state leptons as \cite{Handoko:2001bs};
\begin{equation}
{\cal A}_{LP}^\pm =
\frac{[\Gamma(s_\ell^-,s_\ell^+) + \Gamma(\mp s_\ell^-,\pm s_\ell^+)]
- [\Gamma(\pm s_\ell^-,\mp s_\ell^+) + \Gamma(- s_\ell^-,-
s_\ell^+)]}{[\Gamma(s_\ell^-,s_\ell^+) + \Gamma(\mp s_\ell^-,\pm 
s_\ell^+)] + [\Gamma(\pm s_\ell^-,\mp s_\ell^+) + \Gamma(- s_\ell^-,- 
s_\ell^+)]} . 
\label{eqn:1:4}
\end{equation}

\noindent Using the matrix elment given in eqn.(\ref{eq:1:2}) we can derive the longitudinal polarization asymmetry of $\ell^-$ as;
\begin{eqnarray}
{\cal A}_{LP}^- & = & \frac{ 2 \sqrt{1 - \frac{4 m_{\ell}^2}{m_B^2}}
\mathrm{Re} \left[ \frac{m_B^2}{m_b} C_{Q_1}^* \left( 2 m_{\ell}
C_{10} - \frac{m_{B}^2}{m_b} C_{Q_2} \right) \right] }{\left| 2
m_{\ell} C_{10} + \frac{m_B^2}{m_b} C_{Q_2} \right|^2 + \left( 1
- \frac{4 m_{\ell}^2}{m_B^2} \right) \left| \frac{m_B^2}{m_b} C_{Q_1} 
\right|^2} , 
\label{eq:1:5}
\end{eqnarray}
where ${\cal A}_{LP}^- = {\cal A}_{LP}^+$. Note that within the SM ${\cal A}_{LP} (B_s \to \ell^+ \ell^-)$ is zero. Therefore this asymmetry will be a very useful probe of the new operators in the effective Hamiltonian, in particular to the non CKM-phases in the various Wilson coefficients.


\subsection{The Radiative Decay ($B_s \to \ell^+ \ell^- \gamma$)\label{radiative:1}} 

\par The radiative decay \bstollg can be obtained from the quark level transition \btosll, as given by the effective Hamiltonian in eqn.(\ref{Hamiltonian}). To obtain \bstollg we have to attach a photon line to any of the charged internal and external lines of \btosll. As pointed out in \cite{Aliev:1996ud,Eilam:1996vg} contributions coming from the photon attached to charged internal lines can be neglected. There are two ways to attach the photon line to the external lines; firstly to the external hadronic lines, and secondly to the lepton lines. These shall now be analyzed separately.

\par The task of attaching the photon to the hadronic line can be accomplished by using the following form factors, as defined in Kr\"{u}ger and Melikov \cite{Kruger:2002gf};
\begin{eqnarray}
\langle \gamma (k) | \bar{s} \gamma_{\mu} \gamma_5 b | B(p_B) \rangle
& = & i e \left[ \epsilon_{\mu}^* \left( p_B \cdot k \right) - \left(
p_B \cdot \epsilon^* \right) k_{\mu} \right] \frac{F_A}{m_B} ,
\label{form11}   \\
\langle \gamma (k) | \bar{s} \gamma_{\mu} b | B(p_B) \rangle & = & e
\epsilon^{* \alpha} \epsilon_{\mu \alpha \rho \sigma} p_B^{\rho}
k^{\sigma} \frac{F_V}{m_B} , \label{form12} \\
\langle \gamma (k) | \bar{s} \sigma_{\mu \nu} q^{\nu} \gamma_5 b |
B(p_B) \rangle & = & e \left[ \epsilon_{\mu}^* \left( p_B \cdot k
\right) - \left( p_B \cdot \epsilon^* \right) k_{\mu} \right] F_{TA} ,
\label{form13} \\
\langle \gamma (k) | \bar{s} \sigma_{\mu \nu} q^{\nu} b | B(p_B)
\rangle &=& i e \epsilon^{* \alpha} \epsilon_{\mu \alpha \rho \sigma}
p_B^{\rho} k^{\sigma} F_{TV} , 
\label{form14} 
\end{eqnarray}
where $\epsilon^{\mu}$ and $k^{\mu}$ are the polarization and four momenta of the photon. The definition of the form factors is given in Appendix \ref{appendix:a}. We can express the matrix element for this first decay process as;
\begin{eqnarray}
{\cal M}_1 & = & \displaystyle \frac{\alpha^{3/2} G_F}{\sqrt{2 \pi}}
V_{tb} V_{ts}^* \Bigg[ \left(\bar{\ell} \gamma^{\mu} \ell \right)
\left( A \epsilon^{* \alpha} \epsilon_{\mu \alpha \rho \sigma}
p_B^{\rho} k^{\sigma} + i B \left\{ \epsilon_{\mu}^* \left( p_B \cdot
k \right) - \left( p_B \cdot \epsilon^* \right) \right\} \right)
\nonumber \\  
&& \hspace{1.in} +  \left(\bar{\ell} \gamma^{\mu} \gamma_5 \ell
\right) \left(C \epsilon^{* \alpha} \epsilon_{\mu \alpha \rho \sigma}
p_B^{\rho} k^{\sigma} + i D \left\{ \epsilon_{\mu}^* \left( p_B \cdot
k \right) - \left( p_B \cdot \epsilon^* \right) \right\} \right)
\Bigg] , 
\label{Matrix-elem1}
\end{eqnarray}
in writing the above we have used $e = \displaystyle{2\sqrt{\pi \alpha}}$. And where we have used the following appropriately defined constants;
\begin{eqnarray}
A & = &  2 C_7 \frac{m_b}{q^2} F_{TV} + C_9 \frac{F_V}{m_B} ,
\nonumber \\ 
B & = & - 2 C_7 \frac{m_b}{q^2} F_{TA} - C_9 \frac{F_A}{m_B} ,
\nonumber \\ 
C & = & C_{10} \frac{F_V}{m_B} , \nonumber \\
D & = & - C_{10} \frac{F_A}{m_B} .
\label{constantsM1}
\end{eqnarray}

\par The second process we need to consider is where the photon is attached to the lepton lines. This will mean making the following substitution to the lepton operators in our initial Hamiltonian;
$$ \bar{\ell} {\cal O} \ell \to i e \bar{\ell} \left( \not\!\epsilon^*
\frac{1}{ \not\!p_+ + \not\!k - m_{\ell}} {\cal O} \right) \ell + i e
\bar{\ell} \left( {\cal O} \frac{1}{ - \not\!p_- - \not\!k - m_{\ell}}
\not\!\epsilon^* \right) \ell , $$ 
and using the form factors given in eqn.(\ref{eq:1:1}).

\par We can therefore express the matrix element for this second process as;
\begin{eqnarray}
{\cal M}_2 & = & \displaystyle \frac{\alpha^{3/2} G_F}{\sqrt{2 \pi}}
V_{tb} V_{ts}^* \Bigg[ \left( 2 m_{\ell} C_{10} + \frac{m_B^2}{m_b}
C_{Q_2} \right) \left\{ \bar{\ell} \left( \frac{\not\!\epsilon^*
\not\!p_B}{2 p_+ \cdot k} - \frac{\not\!p_B \not\!\epsilon^*}{2 p_-
\cdot k}\right) \gamma_5 \ell \right\} \nonumber \\ 
&& \hspace{.2in} + \frac{m_B^2}{m_b} C_{Q_1} \left\{ 2 m_{\ell} \left(
\frac{1}{2 p_+ \cdot k} + \frac{1}{2 p_- \cdot k} \right) \bar{\ell} 
\not\!\epsilon \ell + \bar{\ell} \left( \frac{\not\!\epsilon^*
\not\!p_B}{2 p_+ \cdot k} - \frac{\not\!p_B \not\!\epsilon^*}{2 p_-
\cdot k} \right) \ell \right\} \Bigg] .  
\label{Matrix-elem2}
\end{eqnarray}
\noindent The total matrix element would be;
\begin{equation}
{\cal M} = {\cal M}_1 + {\cal M}_2 , \label{tot-mat}
\end{equation}
hence squared matrix element can be written as;
\begin{eqnarray}
|{\cal M}|^2 & = & |{\cal M}_1|^2 + |{\cal M}_2|^2 + 2
\mathrm{Re}\left({\cal M}_1 {\cal M}_2^* \right) ,
\end{eqnarray}
where;
\begin{eqnarray}
|{\cal M}_1|^2 & = & \left|\frac{ \alpha^{3/2} G_F}{\sqrt{2 \pi}}
V_{tb} V_{ts}^* \right|^2 8 \Bigg[  
\left( |A|^2 + |B|^2 \right) \left\{ 2 m_{\ell}^2 \left( (k \cdot
p_+)^2 + (k \cdot p_-)^2 + (k \cdot p_+) (k \cdot p_-) \right)
\right. \nonumber \\ 
&& \hspace{2.2in} \left. + (k \cdot p_-)^2 p_- \cdot p_+ + (k \cdot
p_+)^2 p_- \cdot p_+ \right\} \nonumber \\ 
&& \hspace{.6in} + \left( |C|^2 + |D|^2 \right) \left\{ - 2
m_{\ell}^2 (k \cdot p_-) (k \cdot p_+) + p_- \cdot p_+ \left( (k \cdot
p_+)^2 + (k \cdot p_-)^2 \right) \right\} \nonumber \\ 
&& \hspace{1.2in} + \mathrm{Re}( A^* D + B^* C ) q^2 \left\{ (k \cdot 
p_+)^2 + (k \cdot p_-)^2 \right\} \Bigg] , \label{M1sq}
\end{eqnarray}
\begin{eqnarray}
|{\cal M}_2|^2 & = &  \left|\frac{\alpha^{3/2} G_F}{\sqrt{2 \pi}}
V_{tb} V_{ts}^* \right|^2 16 m_{\ell}^2  f_B^2 ~~ 
\Bigg[ \left| C_{10} + \frac{m_B^2}{2 m_{\ell}
m_b} C_{Q_2} \right|^2 
\left\{ 2 + \frac{ k \cdot p_+ + 2 p_- \cdot p_+ - 
m_{\ell}^2 }{k \cdot p_-} \right. \nonumber \\
&& \left. + \frac{ k \cdot p_- + 2 p_- \cdot p_+ -
m_{\ell}^2 }{k \cdot p_+} + \frac{ q^2 (p_- \cdot p_+ )}{(k \cdot p_-
)(k \cdot p_+)}  - \frac{m_{\ell}^2 ( m_{\ell}^2 + k \cdot p_+
+ p_+ \cdot p_- )}{( k \cdot p_-)^2}  \right. \nonumber \\
&& \left. - \frac{m_{\ell}^2 ( m_{\ell}^2 +
k \cdot p_- + p_+ \cdot p_- )}{( k \cdot p_+)^2} \right\}
+ \left|\frac{m_B^2}{2 m_{\ell} m_b} C_{Q_1}
\right|^2 \left\{ 2 + \frac{ k \cdot p_+ + 2 p_- \cdot p_+ -
m_{\ell}^2 }{k \cdot p_-} \right. \nonumber \\
&& \left. + \frac{ k \cdot p_- + 2 p_- \cdot p_+ - m_{\ell}^2 }{k
\cdot p_+} + \frac{ (p_- \cdot p_+ - m_\ell^2) (p_- \cdot p_+ )}{(k
\cdot p_- )(k \cdot p_+)} + \frac{m_{\ell}^2 ( m_{\ell}^2 - k \cdot
p_+ - p_+ \cdot p_- )}{( k \cdot p_-)^2} \right. \nonumber \\ 
&& \hspace{1in} \left. + \frac{m_{\ell}^2 ( m_{\ell}^2 - k \cdot p_-
- p_+ \cdot p_- )}{( k \cdot p_+)^2} \right\} \Bigg] , \label{M2sq} 
\end{eqnarray}
and
\begin{eqnarray}
2 ~ \mathrm{Re}({\cal M}_1 {\cal M}_2^* ) & = & 
\left|\frac{\alpha^{3/2} G_F}{\sqrt{2 \pi}} V_{tb} V_{ts}^*\right|^2 
16 m_{\ell}^2 f_B ~~ \Bigg[ -
\mathrm{Re} \left\{ \left(C_{10} + \frac{m_B^2 }{2 m_{\ell} m_b}
C_{Q_2} \right) A^* \right\} \times \nonumber \\
&& \frac{( q \cdot k )^3 }{ (p_+ \cdot k )(
p_- \cdot k )} + \mathrm{Re} \left\{ \left( C_{10} + \frac{m_B^2
}{2 m_{\ell} m_b} C_{Q_2} \right) D^* \right\} \frac{ (q \cdot k)^2 (
p_- \cdot k - p_+ \cdot k )}{(p_+ \cdot k )( p_- \cdot k )} 
\nonumber \\ 
&& \hspace{.2in} + \mathrm{Re} \left\{ \left( \frac{m_B^2 }{2
m_{\ell} m_b} C_{Q_1} \right) B^* \right\} \frac{1}{(p_+ \cdot k )(
p_- \cdot k )} ~ \Bigg[ - ( p_- \cdot k )^3   \nonumber \\
&& \hspace{.5in}+ ( p_- \cdot k )^2 ( 2 m_{\ell}^2 - 3 p_+ \cdot k ) 
 - ( p_+ \cdot k ) ( 3 p_- \cdot k + 4 p_- \cdot p_+
) ( p_- \cdot k ) \nonumber \\ 
&& \hspace{.5in} + ( p_+ \cdot k )^2 ( 2 m_{\ell}^2 - p_+ \cdot k
) \Bigg]  \nonumber \\
&& \hspace{.2in} 
+ \mathrm{Re} \left\{ \left( \frac{m_B^2 }{2 m_{\ell}
m_b} C_{Q_1} \right) C^* \right\} \frac{(q \cdot k)^2 ( p_- \cdot k -
p_+ \cdot k )}{(p_+ \cdot k )( p_- \cdot k )} \Bigg] . \label{2ReM1M2}
\end{eqnarray}

\par The differential decay rate is then;
\begin{equation}
\frac{d \Gamma}{d\hat{s}} = \frac{m_B^5}{2^9 \pi^3} 
\left| \frac{\alpha^{3/2} G_F}{\sqrt{2 \pi}} V_{tb} V_{ts}^*
\right|^2  \bigtriangleup ,
\end{equation}
with;
\begin{eqnarray}
\bigtriangleup &=&
\frac{4}{3} m_B^2 ( 1 - \hat{s} )^3 \sqrt{1 - \frac{4
\hat{m}_{\ell}}{\hat{s}}}
\Bigg[ ( \hat{s} + 2 \hat{m}_{\ell}^2 )   
\left\{ |A|^2 + |B|^2 \right\}  + ( \hat{s} - 4 \hat{m}_{\ell}^2 )
\left\{ |C|^2 + |D|^2 \right\} \Bigg] \nonumber \\
&& - 64 \frac{f_B^2}{m_B^2}
 \left| C_{10} + \frac{m_B^2 C_{Q_2}}{2
m_{\ell} m_b} \right|^2 \frac{\hat{m}_{\ell}^2}{(1 - \hat{s})}
\left\{ (4 \hat{m}_{\ell}^2
- \hat{s}^2 - 1) \log (Z) + 2 \hat{s} \sqrt{1 - \frac{4
\hat{m}_{\ell}}{\hat{s}}} \right\} \nonumber \\
&& + 16 \left| \frac{C_{Q_1}}{m_b} \right|^2 
\frac{f_B^2}{(1 - \hat{s})}
\left\{ 2 \hat{s} (4 \hat{m}_{\ell}^2 - 1) \sqrt{1 - \frac{4
\hat{m}_{\ell}}{\hat{s}}} + \left(16 \hat{m}_{\ell}^4 - 4
\hat{m}_{\ell}^2 (1 + 2\hat{s}) + \hat{s}^2 + 1 \right) 
\right. \nonumber \\
&& \left. \hspace{.7in} log(Z)
\right\} 
- 32 \hat{m}_{\ell}^2
(1-\hat{s})^2 f_B
\mathrm{Re} \left\{ A^* \left( C_{10} + \frac{m_B^2 C_{Q_2}}{2 m_{\ell}
m_b} \right) \right\} \nonumber \\
&& \hspace{0.7in} + 16 ~ \frac{m_B \hat{m}_{\ell}}{m_b} ~ f_B ~ 
\mathrm{Re}\left\{ B^* C_{Q_1} \right\}
\left\{ 4 ( \hat{m}_{\ell}^2 + \hat{s} - 1) \log (Z) - 2 \sqrt{1 -
\frac{4 \hat{m}_{\ell}}{\hat{s}}} \hat{s} \right\}  , 
\end{eqnarray}
where $\lambda (a,b,c) = a^2 + b^2 + c^2 - 2 a b - 2 a c - 2 b c$ and $Z = \displaystyle{\frac{1 + \sqrt{1 - \frac{4 \hat{m}_{\ell}^2}{\hat{s}}}} {1 - \sqrt{1 - \frac{4 \hat{m}_{\ell}^2}{\hat{s}}}}}$.

\par Next we compute the FB asymmetry associated with the final state lepton. The definition of the FB asymmetry is;
\begin{equation}
A_{FB}(\hat{s}) = 
\frac{\int^1_0 d cos\theta \frac{d^2 \Gamma}{d\hat{s} cos\theta} -
\int^0_1 d cos\theta \frac{d^2 \Gamma}{d\hat{s} cos\theta}}
{\int^1_0 d cos\theta \frac{d^2 \Gamma}{d\hat{s} cos\theta} +
\int^0_1 d cos\theta \frac{d^2 \Gamma}{d\hat{s} cos\theta}} ,
\end{equation}
where $\theta$ is the angle between $\ell^-$ and $\gamma$ in the dileptonic cm frame. Using this definition of FB asymmetry we arrive at the expression for $B \to \ell^+ \ell^- \gamma$ as;
\begin{eqnarray}
A_{FB} &=& \frac{1}{\bigtriangleup}
\Bigg[2 m_B^2 \hat{s} (1 - \hat{s})^3 
\left(1 - \frac{4 \hat{m}_{\ell}}{\hat{s}}\right) Re\left(A^* D + B^*
C\right)  \nonumber \\
&& + 32 ~ f_B ~ \hat{m}_\ell^2 (1 - \hat{s})^2 Log\left(\frac{4
m_\ell^2}{\hat{s}}\right) Re\left[ D^* \left(C_{10} + \frac{m_B^2}{2
m_\ell m_b} C_{Q_2}\right) \right]   \nonumber \\ 
&& + 16 \frac{m_B}{m_b} \hat{m}_\ell (1 - \hat{s})^2 Log\left(\frac{4
m_\ell^2}{\hat{s}}\right) ~ f_B ~ Re\left[C^* C_{Q_1}\right]  \Bigg] .
\end{eqnarray}


\section{Numerical results and discussion \label{result}}

In this section we will discuss the results of the numerical analysis we have carried out in both the pure dileptonic and radiative decays. For this purpose we have divided this section into two subsections devoted to the numerical analysis of the two different decay modes. The input parameters used for our numerical estimates are given in Appendix \ref{appendix:b}.

\par As we have earlier stated, we are interested in estimating the effects of a large $bsZ$ coupling in dileptonic decays. This sort of enhancement having been proposed by Buras \etal \cite{Buras:2003dj} in order to solve the $B \to \pi \pi$ and $B \to \pi K$ puzzle, where it was proposed that the effective $bsZ$ vertex not only be enhanced in magnitude (by more than twice as much as compared to the SM value) but also receive a large phase, making its effective coupling to be predominantly imaginary. Their fitting effectively makes $C_{10}$ complex, and can be defined as;
\begin{equation}
C_{10} = - (2.2/sin^2\theta_w)e^{i\theta_Y} \quad, \quad \theta_Y =
- (100 \pm 12)^\circ .
\label{enhanced:c10}
\end{equation}
We will be using this value for $C_{10}$ in our analysis with the central value of $\theta_Y$.

\par We have carried out our analysis with the SM using the enhanced value of $C_{10}$ above. We have also used this enhanced value in our numerical work for the 2HDM and SUSY models. Note that both the 2HDM and SUSY models have the universal feature, in connection to the rare decays, that if we are in a large $tan\beta$\footnote{$tan\beta$ is the ratio of the vev's of the two Higgs doublets which are present} region of their parameter space then we have to introduce ``{\sl nonstandard}" operators, namely the scalar and pseudo-scalar operators. In our case this means the introduction of operators corresponding to the Wilsons operators $C_{Q_1}$ and $C_{Q_2}$ in eqn.(\ref{Hamiltonian}).

\par For the case of the 2HDM we have worked in the type-II of this model. The Higgs sector of the type-II 2HDM has almost similar structure and Yukawa couplings to the MSSM. The details of this model have been given in many earlier works \cite{Barger:1989fj,Logan:2000iv}. In the 2HDM type-II we have used the values of the Wilsons as given in \cite{Barger:1989fj,Logan:2000iv}. The experimental observation of $b \to s \gamma$ provides us with a constraint on the charged Higgs mass of this model, that is $m_{H^\pm} \ge 350$ GeV \cite{Gambino:2001ew}.

\par Supersymmetry (SUSY) is one of the most favoured extensions of the SM. But as is well known, even the minimal SUSY extension of the SM, known as MSSM, has a very large number of parameters, making it difficult to do any phenomenology. However, the vast parameter space of the MSSM can be reduced to a manageable level by assuming some unification of the parameters of the MSSM. There are many such models generically named as constrained MSSM (CMSSM) available. Among those one of the very widely used models is the minimal Supergravity (mSUGRA) model. This model has only five parameters, namely; $m$ (the unified mass of all the scalars), $M$ (the unified mass of all gauginos), $A$ (the unified trilinear couplings), $tan\beta$ (the ratio of vev's of the Higgs doublets) and the sign of $\mu$. We have performed our analysis of the dileptonic decays with the mSUGRA model also.


\subsection{Pure dileptonic decay (\bstoll)\label{dilep:2}}

The expressions for the branching ratio and the longitudinal polarization asymmetry for the final state lepton are given in eqns.(\ref{eq:1:3}) and (\ref{eq:1:5}) respectively. Note that Buras \etal \cite{Buras:2003dj} have already noted the enhancement in the branching ratio of \bstomm by using the new $bsZ$ vertex.

\par
\begin{figure}[ht]
\begin{center}
\epsfig{file=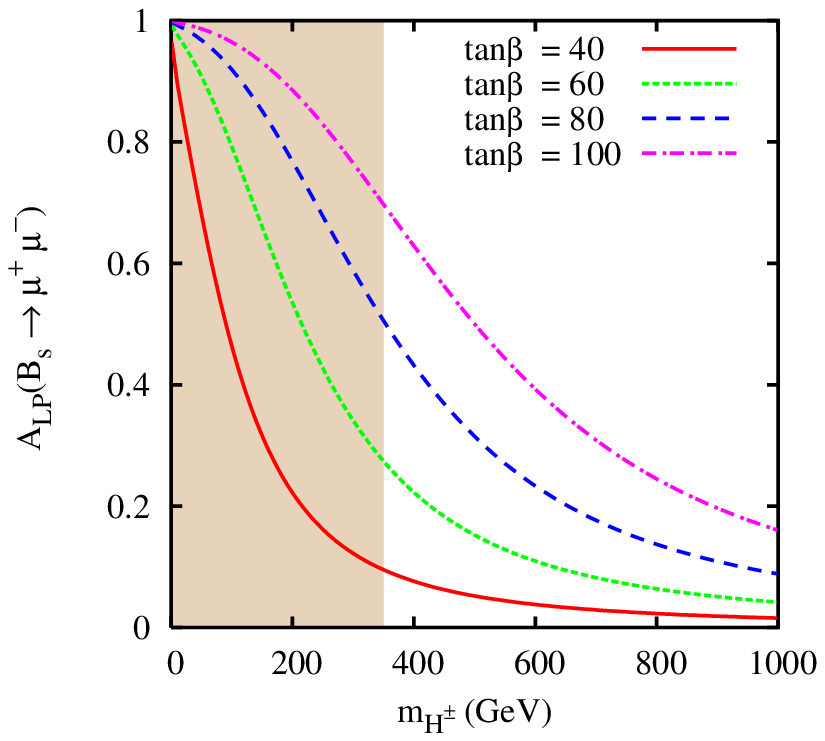,width=.55\textwidth} \hspace{-2cm}
\epsfig{file=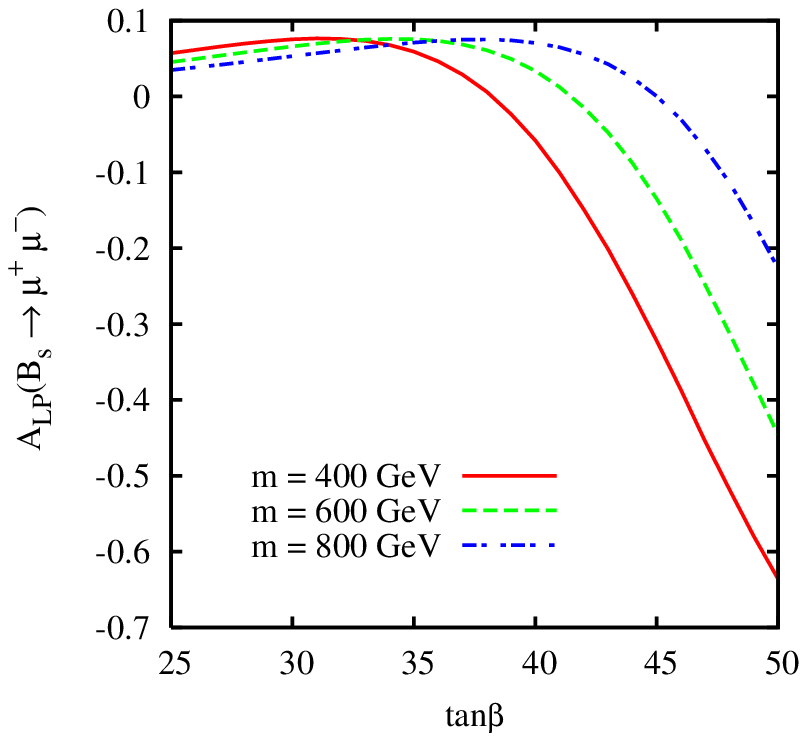,width=.55\textwidth} 
\caption{\it $B_s \to \mu^+ \mu^-$: 2HDM (left) and SUSY (right). The shaded region in the 2HDM figure (left) indicates the region ruled out by the $b \to s \gamma$ observations.}\label{fig:1}\end{center}
\end{figure}

 The enhancement of the branching ratio of the pure dileptonic decay mode (\bstomm) in the presence of the new set of scalar and pseudo-scalar operators has been emphasized earlier \cite{Choudhury:1998ze,Logan:2000iv,Skiba:1992mg,Babu:1999hn,Bobeth:2001sq}. Handoko \etal \cite{Handoko:2001bs} argued that these operators in the dileptonic decay (\bstoll) also provides a non-zero value of the longitudinal polarization asymmetry for the final state lepton. This can also be used as a direct measurement of the physics of scalar and pseudo-scalar interactions. They also presented the correlation between the polarization asymmetry (${\cal A}_{LP}$) and the branching ratio of this mode. At this point we would like to note that this correlation will be modified substantially if we have a large complex phase in the electroweak penguins as suggested by Buras \etal \cite{Buras:2003dj}. Therefore the polarization asymmetry in the dileptonic mode can not only serve as a test of the new set of operators, but it can also provide us with a useful insight in to the nature of the electroweak penguins. To substantiate this point we have presented two plots of ${\cal A}_{LP}$ in figure \ref{fig:1}. The first plot being ${\cal A}_{LP}$ as a function of the charged Higgs boson mass ($m_H^\pm$) for various values of $tan\beta$. In the second plot we have given the same plot in the mSUGRA model for various values of the unified scalar mass ($m$). For both the plots we have taken the value of $C_{10}$ as given in eqn.(\ref{enhanced:c10}). As we can see from these plots there is a marked change in the behaviour of ${\cal A}_{LP}$ as compared to the case where the $bsZ$ vertex is of the SM type. For the set of input parameters we have chosen we calculate the SM branching ratio as $Br(B_s \to \mu^+ \mu^-) = 3.41 \times 10^{-9}$. Using the value of $C_{10}$ as given by eqn.(\ref{enhanced:c10}) we get $Br(B_s \to \mu^+ \mu^-) = 17.21 \times 10^{-9}$. Such that the ratio of the branching ratios is;
$$ \frac{Br(B_s \to \mu^+ \mu^-)}{Br(B_s \to \mu^+ \mu^-)_{SM}} \approx 5 . $$

\subsection{Radiative decay (\bstollg)\label{radiative:2}}

The radiative decay in the SM does not have the helicity suppression of the pure dileptonic decay. This feature makes this decay very useful. Aside from this, in the radiative mode one can measure a large number of observables, such as the FB asymmetry \cite{Kruger:2002gf,Aliev:1996ud,Geng:2000fs,Dincer:2001hu}, and the single and double polarization asymmetries \cite{RaiChoudhury:2002hf} associated with the final state lepton pair. For calculations of the observables in the radiative mode we require the definition of the form factors. In the literature there have been many definitions of the form factors required for the $B \to \gamma$ transition. The FB asymmetry for \bstollg shows a very strong dependence on the specific form of the form factors being used. Where the literature provides us with form factors based on QCD sum rules \cite{Aliev:1996ud}, the quark model \cite{Geng:2000fs} etc. However, as argued by Kr\"{u}ger and Melikov these form factors contradict some of the basic properties. They further gave another parameterization of the form factors, as stated in Appendix \ref{appendix:a:km}. In the plots of the branching ratio and FB asymmetries we have used the form factors given by Kr\"{u}ger and Melikov \cite{Kruger:2002gf}. We have also presented some of the results which one can obtain from the form factors given by Dincer and Sehgal \cite{Dincer:2001hu}. Dincer and Sehgal have used a universal form of the form factors as given in Appendix \ref{appendix:a:ds}. The form factors given by Kr\"{u}ger and Melikov \cite{Kruger:2002gf} and Dincer and Sehgal \cite{Dincer:2001hu} have one common feature, which differs from the earlier formulations of the form factors. The common feature being that away from the charmonium resonances the FB asymmetry vanishes. The zero of the FB asymmetries in \bstollg is a typical feature of the SM, and can be predicted with small theoretical uncertainities. This phenomena is similar to the case of the $B \to K^* \ell^+ \ell^-$ decay where the zero of the FB asymmetry can be used as a tool for the measurement of $C_7^{eff}/Re(C_9^{eff})$ \cite{Beneke:2001at}. Our results also reaffirms the behaviour that FB asymmetry does have a zero in \bstollg.

\begin{figure}[htb]
\begin{center}
\epsfig{file=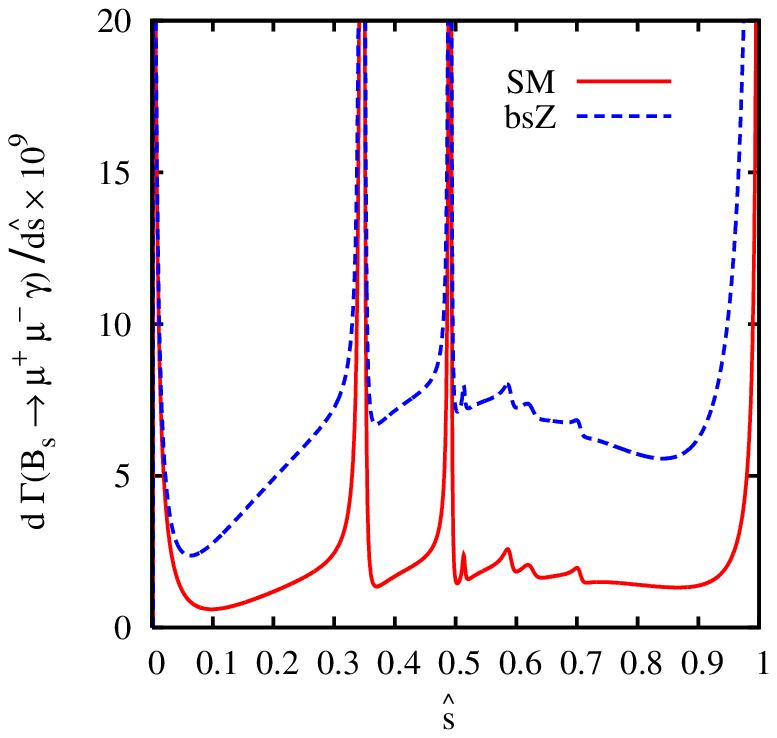,width=.55\textwidth} \hspace{-2cm}
\epsfig{file=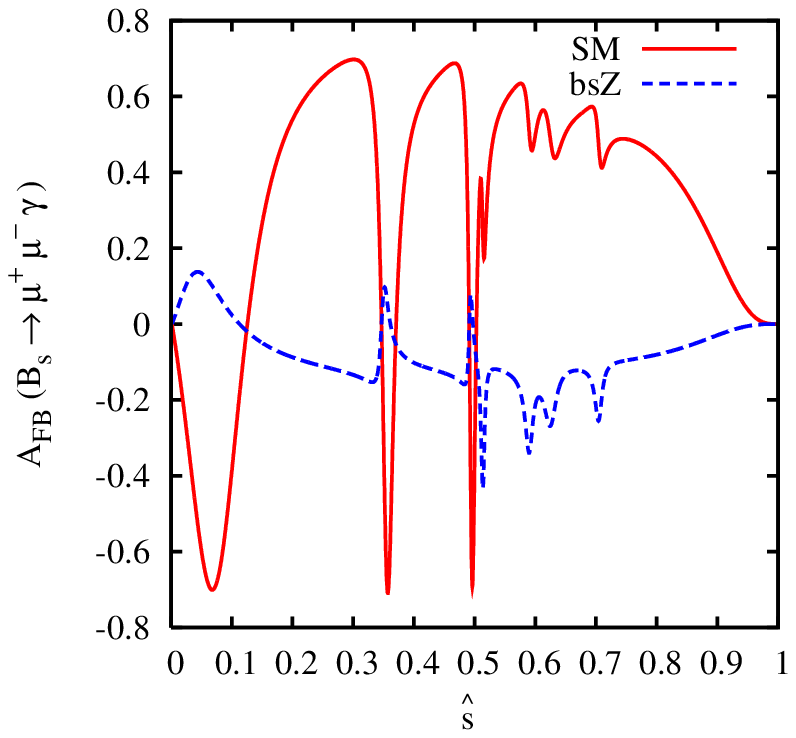,width=.55\textwidth} 
\caption{\it The differential decay rate (left) and FB asymmetry (right) for $B_s \to \mu^+ \mu^- \gamma$.}\label{fig:2} 
\end{center}
\end{figure}

\par In our results we have plotted the branching ratio and FB asymmetries as a function of the dilepton invariant mass. In the set of graphs, following these, we have also shown the results of varying the integrated branching ratio and FB asymmetries against the parameters of the 2HDM and mSUGRA. For calculating the branching ratio and averaged FB asymmetries in the case where $\mu$ is the final state lepton we have excluded the region between $0.33 \le \hat{s} \le 0.55$. This region corresponds to the $J/\psi$ and $\psi'$ resonances. Essentially, in the case of the final state lepton being $\mu$ we have divided the region of the dilepton invariant mass in to two regions, defined as;
\begin{table}[h]
\begin{center}
\begin{tabular}{c c c}
Region - I   & $\Longrightarrow$ &  $4 \hat{m}_\ell^2 \le \hat{s} \le
0.33$  \\  
Region - II  & $\Longrightarrow$ &  $0.55 \le \hat{s} \le 1 - \delta$ .
\end{tabular}
\end{center}
\vskip -.5cm
\end{table}
\par In our analysis we have used a hard photon in the \bstollg decay and have imposed a cut on the photon energy\footnote{When the photon is ``soft" one has to consider both the processes \bstoll and \bstollg together. By taking these two processes together the infrared divergences of \bstollg are cancelled by order $\alpha$ corrections of \bstoll} ($E_\gamma$). Where this photon energy cut can be related to the parameter $\delta$ by the relation $\delta = 2 E_\gamma/m_B$. Our estimates have used a photon energy cut of $E_\gamma = 20$MeV. In the case of the final state lepton being $\tau$ we have used the range $0.55 \le \hat{s} \le 1 - \delta$.

\par
\begin{figure}[ht]
\begin{center}
\epsfig{file=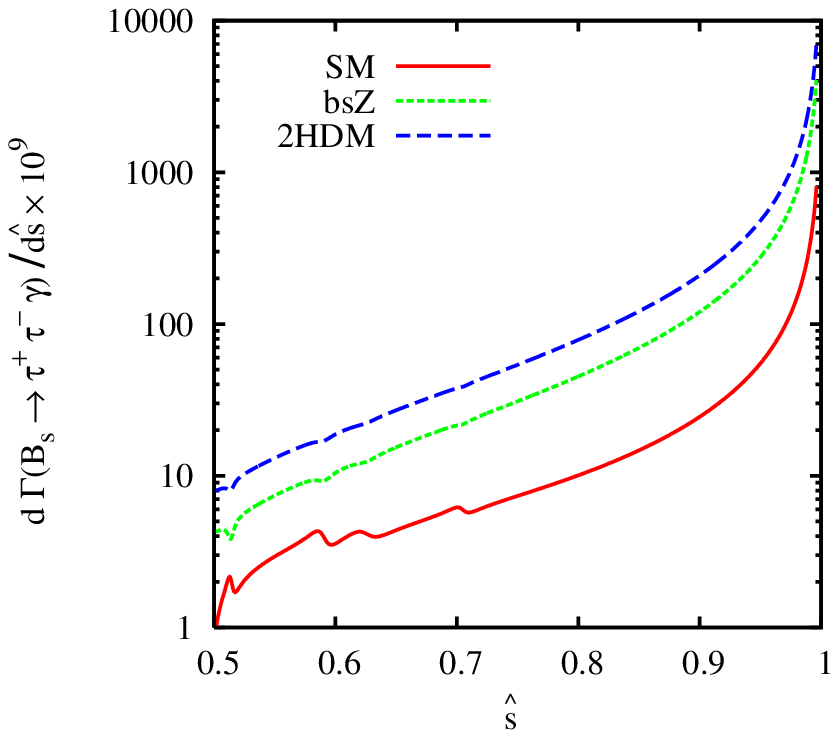,width=.55\textwidth} \hspace{-2cm}
\epsfig{file=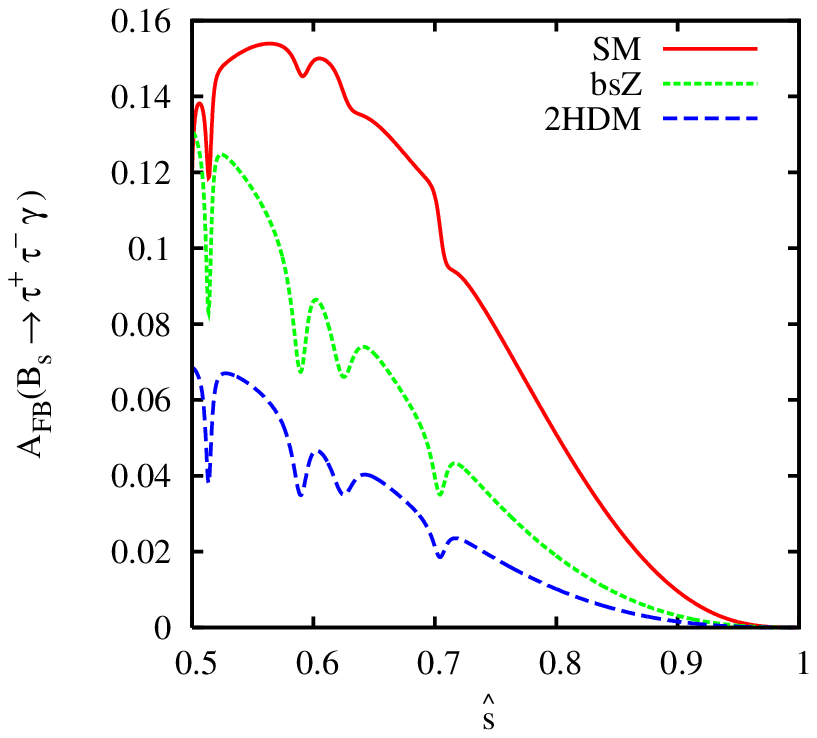,width=.55\textwidth} 
\caption{\it The differential decay rate (left) and FB asymmetry (right) for $B_s \to \tau^+ \tau^- \gamma$ in the 2HDM type-II model. The model parameters are $m_{H^\pm} = 400$GeV and $tan\beta = 80$.}\label{fig:3} 
\end{center}
\end{figure}

 The plots of the variation of the differential decay rate and FB asymmetry as a function of the dilepton invariant mass (for \bstommg) are shown in figure \ref{fig:2}. Similarly the same plot for $\tau$ in the final state is given in figure \ref{fig:3}. As we can see from both figures there can be a substantial variation in both these observables when we have an enhanced $bsZ$ coupling.

\begin{table}[ht]
\begin{center}
\begin{tabular}{| c | c | c | c | c |} \hline
~~Model~~  & ~~Observable~~  & ~~ Region - I ~~ & ~~ Region - II ~~ & 
~~ Total ~~ \\ 
\hline \hline 
SM (K \& M) &  Branching ratio $\times 10^9$ & 0.715  & 1.08 &  1.796
\\ 
       &  $<A_{FB}>$  &  0.265  &  0.30 & 0.286   \\   \hline
bsZ (K \& M)  &  Branching ratio $\times 10^9$ & 1.93 & 4.75 & 6.68
\\ 
       &  $<A_{FB}>$  &  - 0.05  &  -0.07  &  -0.068  \\ \hline \hline
SM  (D \& S) &  Branching ratio $\times 10^9$ & 1.2 & 1.74 & 2.94   \\
       &  $<A_{FB}>$ & 0.3  & 0.45  & 0.39   \\   \hline
bsZ (D \& S)  &  Branching ratio $\times 10^9$ & 3.26 & 7.21 & 10.47
\\ 
       &  $<A_{FB}>$ & -0.071  &  -0.11  & -0.098   \\ \hline
\end{tabular}
\caption{{\it Integrated decay rate and averaged FB asymmetries for \bstommg. ``K \& M'' refers to the results obtained using the form factors of Kr\"{u}ger and Melikov \cite{Kruger:2002gf} whilst ``D \& S'' the results obtained using Dincer and Sehgal's form factors \cite{Dincer:2001hu}. $bsZ$ refers to results obtained using the value of $C_{10}$ as given in eqn.(\ref{enhanced:c10}).} \label{table:1}}
\end{center}
\end{table}
\par We have presented the results of the branching ratio and averaged FB asymmetry for \bstommg in Table \ref{table:1}. In the table we have shown our predictions for the SM and with the $C_{10}$ value given by eqn.(\ref{enhanced:c10}). We have also quoted the results using form factor definitions for both Kr\"{u}ger and Melikov as well as Dincer and Sehgal. As evident from the values given in Table \ref{table:1} the D \& S form factors leds to higher values for the branching ratio and averaged FB asymmetries as compared to K \& M form factors. We can also observe that with the introduction of an enhanced $bsZ$ coupling we have;
$$ \frac{Br(B_s \to \mu^+ \mu^- \gamma)}{Br(B_s \to \mu^+ \mu^- \gamma)_{SM}} \approx 3.5 . $$
We also observe that;
\begin{eqnarray}
\Bigg[\frac{Br(B_s \to \mu^+ \mu^- \gamma)}{Br(B_s \to \mu^+ \mu^-
\gamma)_{SM}}\Bigg]_{4 \hat{m}_\ell^2 \le \hat{s} \le 0.33} 
&\approx & ~~~ 2.7 , \nonumber \\
\Bigg[\frac{Br(B_s \to \mu^+ \mu^- \gamma)}{Br(B_s \to \mu^+ \mu^-
\gamma)_{SM}}\Bigg]_{0.55 \le \hat{s} \le (1 - \delta)} 
&\approx & ~~~ 4.2 , \nonumber 
\end{eqnarray}
which means that the enhancement in the branching ratio is more in the high dilepton invariant mass ($\hat{s}$) region-II. In the case of the FB asymmetries in \bstommg not only does the magnitude decrease but there is also a change in the sign of the averaged FB asymmetries.

\begin{figure}[ht]
\begin{center}
\epsfig{file=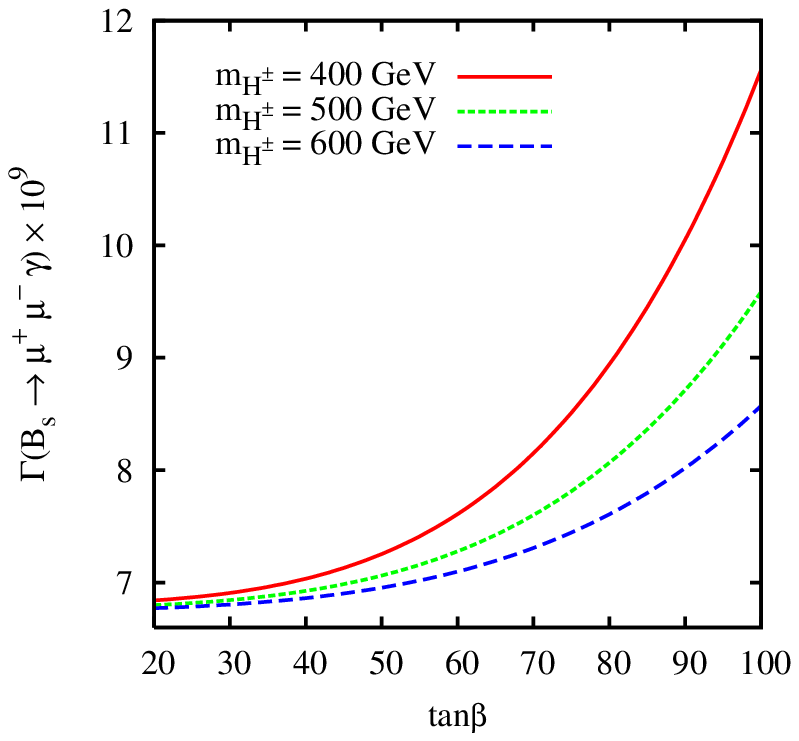,width=.55\textwidth} \hspace{-2cm}
\epsfig{file=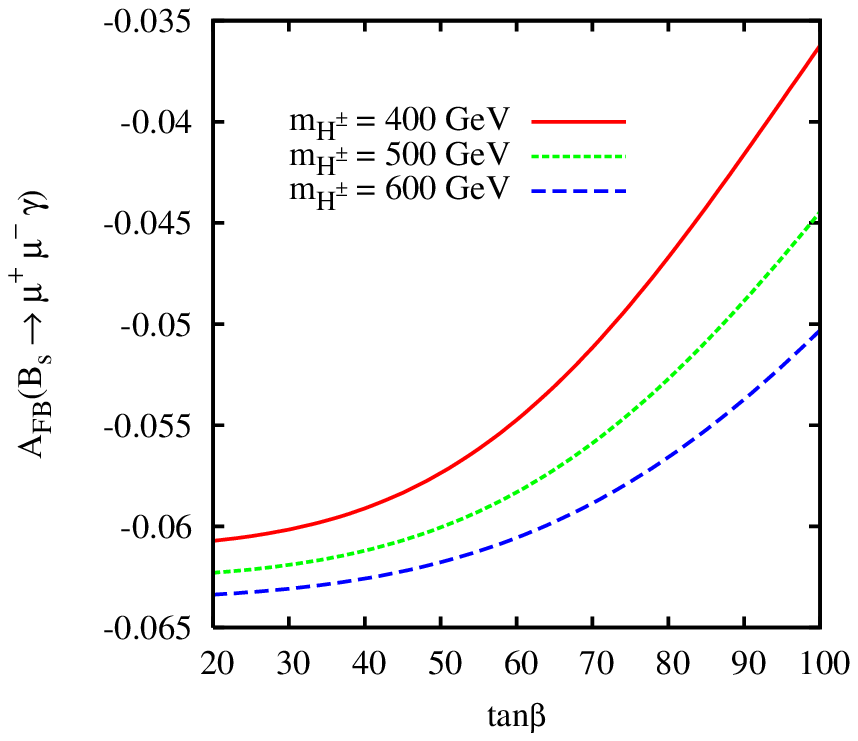,width=.55\textwidth} 
\caption{\it Branching ratio (left) and averaged FB asymmetry (right) for the $B_s \to \mu^+ \mu^- \gamma$ decay in the 2HDM.}\label{fig:4}
\end{center}
\end{figure}
\begin{figure}[ht]
\begin{center}
\epsfig{file=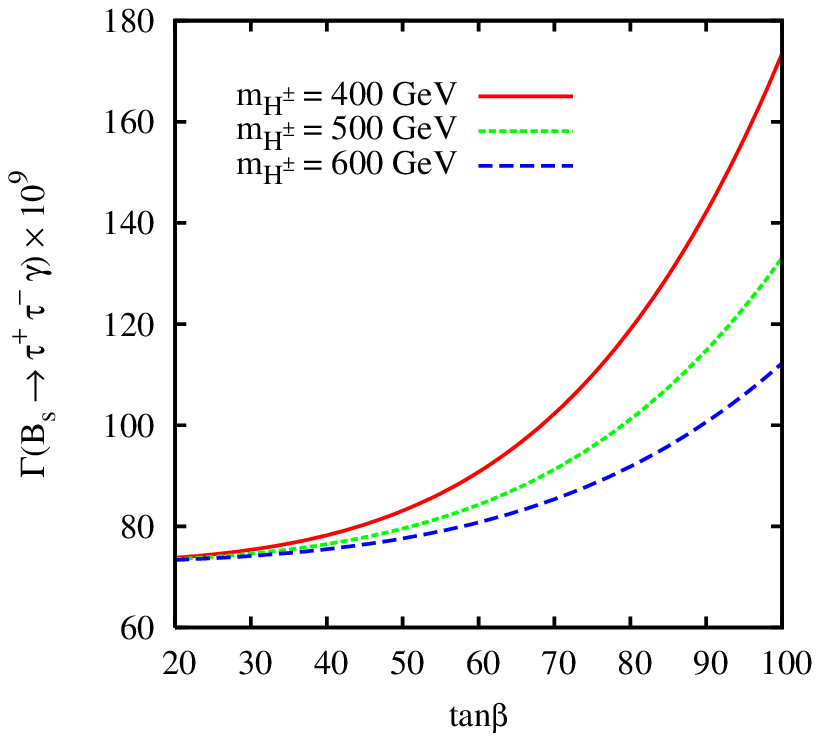,width=.55\textwidth} \hspace{-2cm}
\epsfig{file=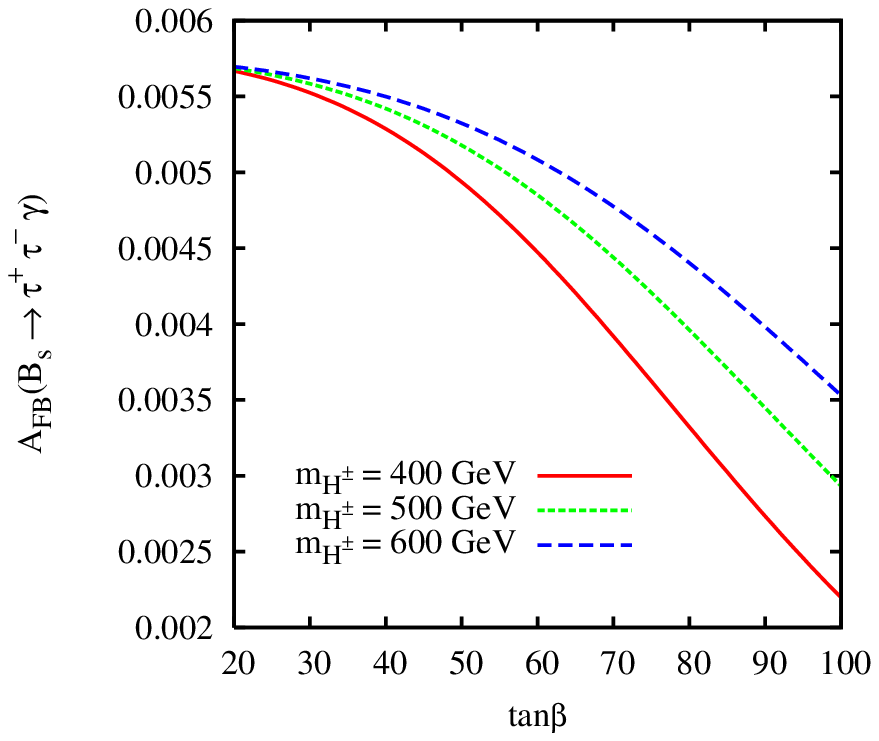,width=.55\textwidth} 
\caption{\it Branching ratio (left) and averaged FB asymmetry (right) for \bstottg in the 2HDM.}\label{fig:5} 
\end{center}
\end{figure}
\begin{table}[ht]
\begin{center}
\begin{tabular}{| c | c | c |} \hline
~~Model~~  & ~~Branching ratio $\times 10^9$~~ & ~~$<A_{FB}> \times
10^{-2}$~~  \\  
\hline \hline
SM     &  15    &  1.6   \\   \hline 
bsZ    &  72.81 &  6   \\  \hline
2HDM   &  118.9 &  0.33   \\  \hline
mSUGRA &  94    &  0.3   \\  \hline
\end{tabular}
\caption{{\it Integrated decay rate and averaged FB asymmetries for \bstottg. In the above results we have used the K \& M form factors \cite{Kruger:2002gf}. The parameters for the 2HDM model are: $m_{H^\pm} = 400$GeV and $tan\beta = 80$. Parameters for the mSUGRA model are: $m = M = 500$GeV, $A = 0$, $tan\beta = 45$ and $sgn(\mu)$ is taken to be positive.}\label{table:2}}
\end{center}
\end{table}
\par In table \ref{table:2} we have quoted the branching ratio and averaged FB asymmetry values for the \bstottg decay in the two models we have considered. We can observe that the enhanced $C_{10}$ Wilson gives an enhancement in the branching ratio of \bstottg by a factor of 5. We can have far greater enhancements in the branching ratio in the 2HDM and mSUGRA models. But in all these cases the FB asymmetry tends to decrease.

\par In figure \ref{fig:4} we have plotted the integrated branching ratio and averaged FB asymmetries of \bstommg in the 2HDM type-II as a function of $tan\beta$ for various values of the charged Higgs mass. It can be observed from the figures that the branching ratio tends to increase with $tan\beta$ whereas the FB asymmetry tends to decrease in magnitude. We have shown similar plots for \bstottg in figure \ref{fig:5} for the 2HDM type-II. As can be seen from this figure there can be an increase in the branching ratio by more than one order of magnitude as compared with its SM value.

\begin{figure}[ht]
\begin{center}
\epsfig{file=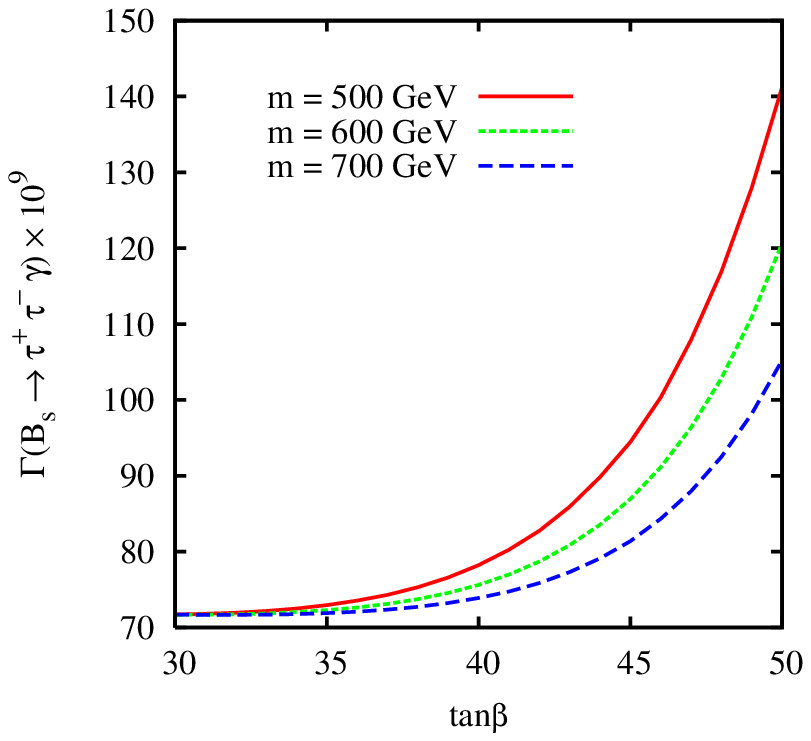,width=.55\textwidth} \hspace{-2cm}
\epsfig{file=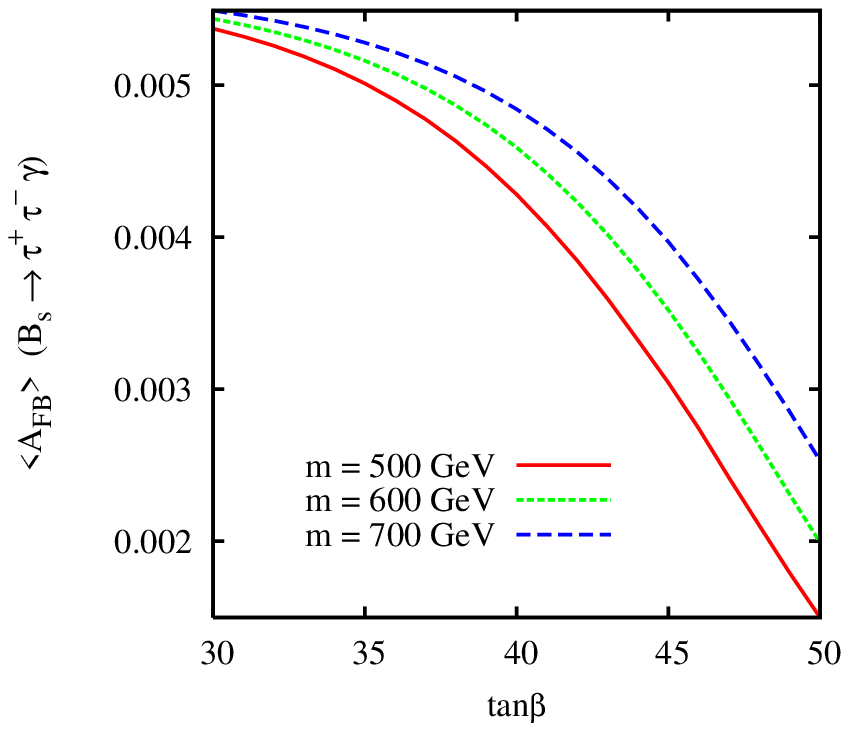,width=.55\textwidth} 
\caption{\it Branching ratio (left) and FB asymmetry (right) for \bstottg in the mSUGRA model. The mSUGRA parameters we have used are $M = 500$GeV, $A = 0$ and $sgn(\mu)$ as positive.}\label{fig:6}
\end{center}
\end{figure}
\par In our final set of graphs, figure \ref{fig:6}, we have plotted the integrated branching ratio and averaged FB asymmetries of \bstottg as a function of $tan\beta$ for various values of the unified scalar masses in the mSUGRA model. Here also we have a behaviour similar to the 2HDM model where the branching ratio increases with $tan\beta$ but the FB asymmetry decreases in magnitude as compared to their respective SM values.

\par Radiative decays within the SM have branching ratios $\sim 10^{-9}$ (for $\mu$) and $\sim 10^{-8}$ (for $\tau$) which can be enhanced by an order in magnitude if some of the new physics comes in to play. The enhanced $bsZ$ vertex not only predicts an increase in the magnitude of the branching ratio but also predicts a sign change in the averaged FB asymmetry in the case of \bstommg relative to the SM. This feature is critically dependent on the nature of the form factors used. Therefore future measurements of \bstollg at hadron collider experiments would provide us with more information on the underlying structure of an effective Hamiltonian which describes \btosll transitions.


\section*{Acknowledgements}
The work of SRC and NG was supported by the Department of Science \& Technology (DST), India, under grant no SP/S2/K-20/99. The work of ASC was supported by the Japan Society for the Promotion of Science (JSPS), under fellowship no P04764.


\appendix


\section{Form Factors \label{appendix:a}}

\subsection{Kr\"{u}ger \& Melikov form factors \label{appendix:a:km}}

The form factors $F_V, F_{TV}, F_A$ and $F_{TA}$ are parameterized as \cite{Kruger:2002gf};
\begin{equation}
F(E_\gamma) = \beta \frac{f_B m_B}{\bigtriangleup + E_\gamma} ,
\end{equation}
where $E_\gamma$ is the photon energy. This energy can be related to the dilepton invariant mass (in the $B$-meson rest frame) as;
$$ E_\gamma = \frac{m_B}{2} \left(1 - \frac{q^2}{m_B^2}\right) . $$ 
For massless leptons the kinematically allowed range would be;
$$0 \le q^2 \le m_B^2 \quad , \quad 0 \le E_\gamma \le E_{max}(= m_B/2) . $$  
For leptons of mass $m_\ell$, the kinematically allowed range for $E_\gamma$ would be;
$$ 4 m_\ell^2 \le q^2 \le m_B^2 \quad, \quad 0 \le E_\gamma \le \frac{m_B^2 - 4 m_\ell^2}{2 m_B} . $$

\begin{table}[h]
\begin{center}
\begin{tabular}{|l|c|c|c|c|}\hline
  & $F_V$ & $F_{TV}$ & $F_A$ & $F_{TA}$ \\ \hline
$\beta$(GeV$^{-1}$) & 0.28 & 0.30 & 0.26 & 0.33 \\ \hline
$\bigtriangleup$ (GeV) & 0.04 & 0.04 & 0.30 & 0.30 \\ \hline
\end{tabular}
\caption{{\it Form factors given by Kr\"{u}ger \& Melikhov \cite{Kruger:2002gf}.}\label{table:3}}
\end{center}
\end{table}

\subsection{Dincer \& Sehgal form factors \label{appendix:a:ds}}

Dincer and Sehgal have used the universal behaviour for all the form factors given by \cite{Dincer:2001hu} in their definition;
\begin{equation}
F_V(E_\gamma) = F_A(E_\gamma) = F_{TV}(E_\gamma) = F_{TA}(E_\gamma)
\approx {1 \over 3} \frac{f_B}{\Lambda_s} \frac{1}{x_\gamma} , 
\label{formfac:ds}
\end{equation}
with $x_\gamma = 2 E_\gamma/m_B$.


\section{Input parameters \label{appendix:b}}
\begin{center}
$m_t = 176$GeV ~~,~~ $m_c = 1.4$GeV ~~,~~ $m_\mu = 0.105$GeV ~~,~~
$m_\tau = 1.77$GeV ,\\ 
$m_B = 5.26$GeV ~~,~~ $m_b = 4.8$GeV ~~,~~ $V_{tb} V_{ts}^* = 0.047$
~~,~~ $\Gamma_B = 4.22 \times 10^{-13}$GeV ,\\ 
$sin^2\theta_w = 0.23$ ~~,~~ $\alpha = 1/130$ ~~,~~ $\Lambda_s =
0.5$GeV. ~~,~~ $f_B = 0.2$GeV.  
\end{center}



\end{document}